\documentclass[floatfix,pra,twocolumn,showpacs,preprintnumbers,amsmath,amssymb,letterpaper]{revtex4}
\usepackage{times}
\usepackage{hyperref}
\usepackage{soul,xspace,braket}
\usepackage{graphicx}
\usepackage{hypernat}

\hypersetup{colorlinks=true,
linkcolor=blue,
citecolor=blue,
urlcolor=blue}

\mathchardef\ordinarycolon\mathcode`\:
\mathcode`\:=\string"8000
\begingroup \catcode`\:=\active
  \gdef:{\mathrel{\mathop\ordinarycolon}}
\endgroup

\newcommand{\di}{\ensuremath{\mathrm{d}}}
\newcommand{\im}{\ensuremath{\mathrm{i}}}
\newcommand{\eu}{\ensuremath{\mathrm{e}}}

\newcommand{\tr}{\ensuremath{\mathrm{tr}}}

\begin{document}
\title{Parameter estimation with cluster states}
\author{Matthias Rosenkranz}
\email{m.rosenkranz@physics.ox.ac.uk}
\author{Dieter Jaksch}
\homepage{http://www.physics.ox.ac.uk/qubit/}
\affiliation{Clarendon Laboratory, University of Oxford, Parks Road,
  Oxford OX1 3PU, United Kingdom}
\affiliation{Keble College, Parks
  Road, Oxford OX1 3PG, United Kingdom}
\pacs{03.65.Ta, 03.67.-a, 06.20.Dk} \date{\today}

\begin{abstract}
  We propose a scheme for parameter estimation with cluster states. We
  find that phase estimation with cluster states under a many-body
  Hamiltonian and separable measurements leads to a precision at the
  Heisenberg limit.  As noise models we study the dephasing,
  depolarizing, and pure damping channels.  Decoherence reduces the
  sensitivity but our scheme remains superior over several reference
  schemes with states such as maximally entangled states and product
  states.  For small cluster states and fixed evolution times it
  remains at the Heisenberg limit for approximately 2 times as many
  qubits than alternative schemes.
\end{abstract}
\maketitle

\section{Introduction}\label{sec:intro}
The precise determination of system parameters by measurements is the
basis of many applications in physics and beyond. Quantum mechanics
offers a way to enhance measurement sensitivity by lowering the
theoretically achievable limits. It is a long-standing goal in the
quantum parameter estimation field to find feasible ways to reach this
so-called Heisenberg limit. We call the number of physical systems
used to measure the parameter (e.g., the number of atoms or photons)
the \emph{resource size}. In the Heisenberg limit, the measurement
precision scales with the inverse resource size. In contrast, the
classical shot-noise limit scales only with the inverse square root of
the resource size. Examples of emerging quantum technology
applications based on this enhancement are improved frequency
standards~\cite{BolItaWin96}, the construction of quantum clocks and
their synchronization~\cite{GioLloMac01}, the detection of weak
forces~\cite{NemMunMil03}, or the improved resolution and
signal-to-noise ratio in image
reconstruction~\cite{GioLloMac09,Llo08}. A straightforward proposal to
achieve the Heisenberg limit in principle by using maximally entangled
states faces the problem that these states are also more sensitive to
background noise and hence decoherence. In general this leads to a
diminishing or complete cancellation of the overall improvement of
these schemes~\cite{HueMacPel97}. In this paper, we present a scheme
based on cluster states, which are a class of entangled states. They
are well known in the context of quantum computation~\cite{RauBri01}
but, to our knowledge, their relevance in parameter estimation theory
has not yet been investigated. Cluster states prove very robust
against many sources of noise~\cite{BriRau01,DueBri04}. We combine
this robustness with a scheme capable of estimating a phase parameter
of the system evolution.

The main results of our proposal can be summarized as (i) cluster
states attain the Heisenberg limit for a setup with a three-body
Hamiltonian and no decoherence.  In atomic systems, cluster states
have been implemented with neutral atoms in an optical
lattice~\cite{ManGreWid03}. Also many other proposals on how to create
cluster states exist, for example, with ions~\cite{IvaVitPle08}, in
cavities~\cite{ZouMat05,ChoLee05}, in charge and flux
qubits~\cite{TanLiuFuj06}, and with the help of a linear quantum
register~\cite{ClaAlvJak05,ClaKleBru07}.  Proposals for implementing
Hamiltonians of the form considered in this paper have been put
forward for neutral atoms in optical lattices~\cite{BueMicZol07} and
cold polar molecules~\cite{PacPle04}.  It should be noted that recent
works have shown that the Heisenberg limit is not the ultimate lower
bound in quantum metrology, but that it can be beaten by implementing
Hamiltonians with symmetric $k$-body
interactions~\cite{BoiFlaCav07,BoiDatFla08,ChoSun08}.  (ii) All
required measurements are of a simple tensor product form involving
only single Pauli measurements. (iii) Decoherence leads to a decrease
in measurement precision but our scheme remains superior to various
reference schemes. In implementations with cold atoms, the main
contribution to decoherence is dephasing noise, which we study in
detail in this paper.  Furthermore, we consider depolarization and
damping noise, which represent smaller contributions to the overall
decoherence in typical realizations.  Our first reference system is
given by a standard quantum metrology scheme with different
Hamiltonian and initial states than the cluster state scheme.  This
will show that our scheme improves on the precision of these standard
schemes.  The second reference system consists of the Hamiltonian of
the cluster state scheme but with different initial states. This will
demonstrate that one can only expect this improvement for a suitable
combination of Hamiltonian and initial state.

Cluster states have been implemented in purely optical setups using
polarization and momentum entanglement of photons currently for up to
six qubits
\cite{WalResRud05,KieSchWeb05,ZhaLuZho06,LuZhoGue07,ValPomMal07}.  The
measurements required for our scheme can also be realized optically.
We are not aware of proposals to implement the Hamiltonian of our
proposed scheme in an optical system and developing such a proposal is
beyond the scope of the present paper.  Decoherence in the optical
case is mainly caused by photon loss, which is not discussed
here~\cite{DorDemSmi09}.

This paper is organized as follows. After the introduction of
parameter estimation theory and cluster states, we discuss in
Sec.~\ref{sec:parameter-estimation} how cluster states can be employed
in parameter estimation schemes. It turns out that, in principle, it
is possible to achieve the Heisenberg limit with this cluster state
scheme. We also define our reference systems. In
Sec.~\ref{sec:estimation-noise}, we show how dephasing noise
influences the precision of the measurements. We discuss an analytical
model of the cluster state measurement scheme and present our main
numerical results. These results are compared with an alternative
estimation scheme and we find an overall improvement of precision
even under noise. Two further noise channels, namely the depolarizing
and damping channels, are considered in Sec.~\ref{sec:depolarizing},
with qualitatively similar results.  In Sec.~\ref{sec:system-II}, we
consider a further reference system subjected to all three noise
channels. The scaling with the number of resources when we include
noise is discussed in Sec.~\ref{sec:scaling}. There we show that our
scheme offers advantages for small resource sizes.  We conclude the paper
in Sec.~\ref{sec:conclusion} and explicitly derive an analytical
solution for our model with resource size $N=2$ in an appendix.

\section{Parameter estimation schemes}\label{sec:parameter-estimation}
In this section, we introduce the basics of parameter estimation theory
and cluster states. Subsequently, we combine the two and introduce a
scheme of parameter estimation with cluster states. Finally, we
present the reference systems we will be using throughout the paper.

\subsection{Parameter estimation theory}\label{ssec:estimation-theory}
We consider a system with Hamiltonian $H_\chi = \chi H_0$ and density
operator $\rho(t)$ which evolves according to the von Neumann equation
($\hbar = 1$)
\begin{equation}\label{eq:Heisenberg-eq}
  \frac{\di\rho}{\di t}(t) = \im [\rho(t), H_\chi].
\end{equation}
Parameter estimation aims to determine the value of the parameter
$\chi$ by means of suitable measurements on many copies of the system.
Repeated measurements of an operator $O$ result in $\nu$ data points
$o_i$. It is then possible to extract an estimate for $\chi$ from this
data by means of an estimator $\chi_\mathrm{est}(o_1, \dots,
o_\nu)$. The uncertainty of the estimation is expressed as
\begin{equation}\label{eq:uncertainty}
  \delta \chi = \left\langle\left( \frac{\chi_\mathrm{est}}{|\di\langle
        \chi_\mathrm{est}\rangle/\di \chi|} - \chi \right)^2
  \right\rangle^{1/2}.
\end{equation}
The denominator cancels superfluous ``units'' introduced by the
estimator. If we assume an unbiased estimator, i.e., $\langle
\chi_\mathrm{est}\rangle = \chi$, then the uncertainty,
Eq.~\eqref{eq:uncertainty}, reduces to the deviation of the estimator,
which is a familiar measure of precision.

If we let $\rho(t)$ be a pure state, then it can be shown that the
uncertainty $\delta \chi$ is bounded
by~\cite{BraCav94,GioLloMac06,BraCavMil96,BoiDatFla08}
\begin{equation}\label{eq:delta-omega}
  \delta \chi \geq \frac{1}{2\sqrt\nu t\Delta H_0}.
\end{equation}
$\Delta H_0$ is the standard deviation of $H_0$, i.e., $\Delta H_0 =
\sqrt{\langle H_0^2\rangle - \langle H_0\rangle^2}$. The bound
Eq.~\eqref{eq:delta-omega} is also called quantum Cram{\' e}r-Rao
bound. There are two routes available to minimizing the
uncertainty. The factor $\sqrt\nu$ in the denominator of
Eq.~\eqref{eq:delta-omega} is the well-known statistical improvement
with the number of independent runs $\nu$. In general, this
$1/\sqrt\nu$ scaling is attained in the limit of many runs of the
experiment~\cite{BraCavMil96}. Here we are not interested in this
trivial improvement. The second and more interesting way is to
maximize $\Delta H_0$.

Consider $N$ replicas of a physical system evolving according to the
Hamiltonian $H_\chi=\chi H_0$ with
\begin{equation}\label{eq:H}
  H_0 = \sum_{j=1}^N h_0^{(j)},
\end{equation}
where $h_0^{(j)}$ is the Hamiltonian of the $j$th replica. We assume
that each replica undergoes the same evolution so that each
$h_0^{(j)}$ is of the same form. The maximal deviation of the
composite Hamiltonian $H_0$ is then attained by evolving the maximally
entangled state
\begin{equation*}
  \ket{\psi} = \frac{1}{\sqrt 2} \left(
    \ket{\lambda_\text{min}}^{\otimes N} + \ket{\lambda_\text{min}}^{\otimes N}
  \right),
\end{equation*}
where $\ket{\lambda_\text{min}}$ and $\ket{\lambda_\text{max}}$ are
the eigenvectors of $h_0^{(j)}$ with the smallest and largest
eigenvalues $\lambda_\text{min}$ and $\lambda_\text{max}$,
respectively. This state yields $\Delta H_0 = N(\lambda_\text{max} -
\lambda_\text{min})/2$, and we see that the generalized uncertainty,
Eq.~\eqref{eq:delta-omega}, is given by
\begin{equation*}
  \delta \chi \geq \frac{1}{\sqrt\nu t N(\lambda_\text{max} -
    \lambda_\text{min})}.
\end{equation*}
Such a scaling with the inverse of the resource size, here $N$, is
called Heisenberg scaling and the corresponding precision measurement
is said to attain the Heisenberg limit. Note that this limit can be
reached by using only separable measurements on each constituent
\cite{GioLloMac06}. In contrast, if we use uncorrelated states
initially, the scaling is proportional to $1/\sqrt N$, which is known
as the standard quantum limit. Quantum metrology offers an improvement
of $1/\sqrt N$ over classical schemes for Hamiltonians of the form in
Eq.~\eqref{eq:H}. However, in realistic setups the von Neumann
equation is rarely exactly realized owing to the presence of
environmental noise.  Intuitively one would expect noise to
deteriorate the precision. In fact, Huelga \textit{et
  al.}~\cite{HueMacPel97} and others~\cite{UlaKit01,ShaCav07} found
that decoherence rapidly destroys the improvement gained in a
straightforward implementation with maximally entangled states. They
also point out the existence of partially entangled states which do
improve the precision even under noise. This paper aims to improve on
this result by introducing the more robust cluster states into quantum
metrology.

\subsection{Cluster states}\label{ssec:cluster}
Cluster states are a particular type of many-body entangled states
\cite{BriRau01,HeiEisBri04}. Their mathematical description is based
on the notion of graphs. A graph $G$ consists of a finite nonempty set
$V$ of $N$ vertices together with a finite set $E$ of $m$ unordered
pairs of distinct vertices from $V$~\cite{Har69}. Elements of $E$ are
also called edges since they join two vertices. Based on this notion
we define the family of operators
\begin{equation*}
  K^{(i)} = \sigma_x^{(i)}\prod_{j \in N(i)} \sigma_z^{(j)},\quad i \in V,
\end{equation*}
where $N(i)$ denotes the set of vertices sharing an edge with the
$i$th vertex, and a Pauli operator $\sigma_{x,y,z}^{(i)}$ acts only on
the $i$th vertex. A graph state $\ket{+_G}$ is defined as the unique
eigenstate with $K^{(i)}\ket{+_G} = \ket{+_G}$ for all operators
$K^{(i)}$.  The group $\mathcal{S}_G$ generated under multiplication
by the set $\set{K^{(i)} | i \in V}$ is called the stabilizer of the
graph state. It is instructive to identify the vertices with a
two-level physical system, which we will henceforth denote as
qubit. With this notation, an equivalent way of describing graph
states is given by first initialising every qubit in the superposition
$(1/\sqrt 2)(\ket 0 + \ket 1)$ and then applying a controlled phase
gate between each pair of qubits connected by an edge in the
graph~\cite{HeiEisBri04}.

For simplicity, in this paper, we reduce the class of all valid graphs
to linear graphs, which results in one-dimensional (1D) cluster
states. Here, all vertices are connected to exactly two neighbors,
which are pairwise distinct (except the first and last vertex, which
are connected to one neighbor each). In physical terms one could
imagine a string of atoms, for example, each interacting with its next
neighbors only. This leads to the family of stabilizers
\begin{equation}\label{eq:stabilizer}
  K^{(i)} = \sigma_z^{(i-1)} \sigma_x^{(i)} \sigma_z^{(i+1)}
\end{equation}
for $1 \leq i \leq N$, where $N$ is the number of vertices (or,
equivalently, qubits). We use the convention that $\sigma_z^{(0)} =
\sigma_z^{(N+1)} = 1$.  Cluster states are known to be remarkably
stable against a range of sources of
noise~\cite{DueBri04,HeiDueBri05,BriRau01,SimKem02}. This stability
can be traced to the fact that local decoherence on an individual
qubit in the composite state only affects the small set of its
neighbors. For cluster states, the number of neighbors does not depend
on the total number of qubits but is a constant of the underlying
graph. In this work we will exploit this feature to achieve a higher
measurement precision even under the influence of noise.

\subsection{Parameter estimation scheme with cluster states}
We have seen in Sec.~\ref{ssec:estimation-theory} that superpositions
of states with maximally separated eigenvalues of $H_0$ together with
appropriate measurements are sufficient to reach the Heisenberg
limit. In Sec.~\ref{ssec:cluster} we have introduced 1D cluster states
as eigenstates of a family of correlators $K^{(i)}$. In analogy to the
simple case of independent one-body Hamiltonians for a system of $N$
qubits, we sum the correlators of all vertices and interpret this sum
as a Hamiltonian
\begin{equation}\label{eq:H-cluster}
  H_{II} = \frac{\chi}{2}\sum_{j=1}^N K^{(j)} = \frac{\chi}{2}\sum_{j=1}^N
  \sigma_z^{(j-1)} \sigma_x^{(j)} \sigma_z^{(j+1)}.
\end{equation}
Here, $\chi$ is the parameter we would like to estimate and $j$ counts
the qubits.  In an atomic optical lattice implementation of this
Hamiltonian~\cite{PacPle04}, the parameter $\chi$ would sensitively
depend on the atomic scattering properties.  In a purely optical
setup, measuring $\chi$ could reveal the values of high-order terms in
the susceptibility of the medium with high precision.  Let $\ket{+_N}$
be a 1D cluster state of $N$ qubits so that $K^{(i)} \ket{+_N} =
\ket{+_N}$ for all $i$. It is straightforward to see that the state
$\ket{-_N} := \prod_{j=1}^N \sigma_z^{(j)}\ket{+_N}$ is an eigenstate
with $K^{(i)}\ket{-_N} = -\ket{-_N}$ for all $i$. To prove this note
that $\sigma_z^{(j)}$ commutes with $K^{(i)}$ for $i\neq j$ but
anticommutes for $i=j$~\cite{HeiDueEis05}. This is analogous to the
eigenstates $\ket{0}$ and $\ket{1}$ of the single particle
Hamiltonians $h_0^{(i)} = \sigma_z^{(i)}$ in conventional quantum
metrology schemes~\cite{HueMacPel97}. We use a superposition of
eigenstates, cluster states, of the underlying Hamiltonian as a
resource, namely,
\begin{equation}\label{eq:psi-cluster}
  \ket{\psi_\text{c}} = \frac{1}{\sqrt 2} (\ket{+_N} + \ket{-_N}).
\end{equation}
Here and in the following, an index c on a quantity denotes that this
quantity belongs to the cluster state setup.  Subsequently, the system
evolves under $H_{II}$ for a time $t$ before it is measured. The time
evolution leaves the system in the state
\begin{equation*}
  \ket{\psi_\text{c}}_t = \frac{1}{\sqrt 2} \left( \eu^{-\im N\chi
      t/2}\ket{+_N} + \eu^{\im N\chi t/2}\ket{-_N} \right).
\end{equation*}
As measurement operator we employ
\begin{equation}\label{eq:M-cluster}
  M_\text{c} = \prod_{j=1}^N \sigma_z^{(j)}.
\end{equation}
The expectation value and deviation of $M_\text{c}$ are
\begin{align}
  \langle M_\text{c}\rangle &= \cos(N\chi t),\label{eq:exp-M}\\
  \Delta M_\text{c} &= |\sin(N\chi t)|.\label{eq:delta-M}
\end{align}
We now consider $\nu$ independent runs of the experiment. In order to
calculate the uncertainty, Eq.~\eqref{eq:uncertainty}, inherent in this
setup, we define an estimator implicitly via
\begin{equation}\label{eq:estimator}
  \frac{1}{\nu}\sum_{j=1}^\nu M_\text{c}^{(j)} =: \cos(N\chi_\text{est} t),
\end{equation}
where $M_\text{c}^{(j)}$ is the measurement operator for the $j$th
independent run. This estimator is unbiased, which can be seen by
expanding the right-hand side of Eq.~\eqref{eq:estimator} around the
actual coupling $\chi$ up to first order and taking the expectation
value combined with Eq.~\eqref{eq:exp-M}.  This reduces the
uncertainty, Eq.~\eqref{eq:uncertainty}, to the deviation
\begin{equation*}
  \delta \chi_\text{c} = \Delta \chi_{\text{est}} = \frac{\Delta
    M_\text{c}}{\sqrt\nu N t |\sin(N\chi t)|} = \frac{1}{\sqrt\nu t N}.
\end{equation*}
Hence, our setup achieves the Heisenberg scaling $1/N$ in the
decoherence-free case. In this sense, the measurement operator
$M_\text{c}$ we chose is optimal for this setup. The factor
$1/\sqrt\nu$ originates from statistical averaging over $\nu$
measurements. The $1/\sqrt N$ improvement over the shot-noise limit is
the result of the $N$-fold increase of the frequency in the
expectation value of $M_\text{c}$.

As with all phase measurement schemes, the periodicity of the
expectation value, Eq.~\eqref{eq:exp-M}, and estimator,
Eq.~\eqref{eq:estimator}, result in an ambiguity when globally
determining the value of $\chi$ without prior knowledge.  However, we
expect our setup to be a good candidate for local phase determination
when the initial value of the parameter is known with a good
accuracy~\cite{DurDow07}.

\subsection{Reference systems}
In the following sections, we compare the attainable precision of our
cluster state setup with different reference systems. The cluster
state setup is defined by the Hamiltonian $H_\chi = H_{II}$,
Eq.~\eqref{eq:H-cluster}, the initial state $\ket{\psi_\text{c}}$,
Eq.~\eqref{eq:psi-cluster}, and the measurement $M_\text{c}$,
Eq.~\eqref{eq:M-cluster}. This setup is always kept invariant in these
comparisons.

\subsubsection{Reference system (I)}
The first reference system is the standard system used in parameter
estimation.  It is defined by the Hamiltonian
\begin{equation*}
  H_I = \frac{\chi}{2} \sum_{j=1}^N \sigma_z^{(j)}.
\end{equation*}
We then use two different initial states, which evolve under this
Hamiltonian, and corresponding measurement operators. These are given
by
\begin{align}
  \ket{\psi_\text{m}} &= \frac{1}{\sqrt{2}} \left(\ket{0}^{\otimes N}
    + \ket{1}^{\otimes N} \right), & M_\text{m} &= \prod_{j=1}^N
  \sigma_x^{(j)},\label{eq:sysI-max}\\
  \ket{\psi_\text{u}} &= \left( \frac{1}{\sqrt{2}} (\ket{0} + \ket{1})
  \right)^{\otimes N}, & M_\text{u} &= \sum_{j=1}^N
  \sigma_x^{(j)}\label{eq:sysI-unc}.
\end{align}
The indices m and u indicate quantities corresponding to maximally
entangled and uncorrelated states, respectively. The measurement
operators $M_\text{m}$ and $M_\text{u}$ are optimal for attaining the
lowest deviation in the decoherence-free case with the respective
initial state~\cite{GioLloMac06}.  If we include dephasing noise in
this setup, we recover the result in~\cite{HueMacPel97} that maximally
entangled states lead to the same minimal deviation $\delta\chi$ as
uncorrelated states. The origin for this behavior is the $N$-fold
increase in the decoherence rate for maximally entangled state
compared to uncorrelated states.

\subsubsection{Reference system (II)}
The Hamiltonian for our second reference system is given by $H_{II}$,
Eq.~\eqref{eq:H-cluster}. As in reference system (I) we use maximally
entangled and uncorrelated states as initial states for the parameter
estimation. This means that we can compare cluster state parameter
estimation with parameter estimation with different initial states
under the same Hamiltonian $H_{II}$.  Also the measurement operators
for the respective states are the same as before.  Hence, reference
system (II) is defined by Hamiltonian $H_{II}$,
Eq.~\eqref{eq:H-cluster}, together with Eqs.~\eqref{eq:sysI-max} and
\eqref{eq:sysI-unc} as initial states and measurements.

\section{Parameter estimation under dephasing
noise\label{sec:estimation-noise}}
Physical systems are often exposed to various sources of noise. In
this section, we focus on individual dephasing of the qubits.
Dephasing is modeled by replacing the unitary evolution
Eq.~\eqref{eq:Heisenberg-eq} with the master equation
\begin{equation}\label{eq:master}
  \frac{\di\rho}{\di t}(t) = \im[\rho(t), H_\chi] + \frac{\gamma}{2}\sum_{j=1}^N
  \left[\sigma_z^{(j)} \rho(t) \sigma_z^{(j)} - \rho(t)\right].
\end{equation}
In this paper, we choose the Hamiltonians $H_I$ or $H_{II}$ for
$H_\chi$ depending on the reference system we study.  The parameter
$\gamma$ denotes the strength of the dephasing. To achieve the Cram{\'
  e}r-Rao bound it is necessary to repeat the single measurements many
times. As in~\cite{HueMacPel97}, we assume a total run time $T$ of the
whole experiment with the duration $t$ of each individual realization.
This results in the total number of experiments $\nu = T/t$.

\subsection{Analytical solution for $N=2$\label{ssec:analytical}}
In this section we discuss an analytical solution of the cluster state
estimation problem with the master equation~\eqref{eq:master} for
the case $N=2$. In the Appendix we present the derivation of this
explicit solution. For higher $N$ our direct approach would also yield
analytical solutions, however, the complexity of the calculation
rapidly becomes intractable. For $N=2$, the expectation value and
deviation of $M_\text{c}$ are given by
\begin{align}
  \langle M_\text{c}\rangle(t) &= \eu^{-\gamma t}\left( \cos(\Omega t)
    +
    \frac{\gamma}{\Omega} \sin(\Omega t) \right),\\
  \delta\chi_\text{c}(t) &= \frac{\eu^{\gamma t} \sqrt{1 -
      \eu^{-2\gamma t} \left( \cos(\Omega t) +
        \frac{\gamma}{\Omega}\sin(\Omega t) \right)^2}}{\sqrt{Tt}
    \frac{4\chi}{\Omega} \left| \frac{\gamma}{\Omega} \cos(\Omega t) -
      \left(1+\frac{\gamma}{\Omega^2 t} \right) \sin(\Omega t)
    \right|},\label{eq:delta_chi_2}
\end{align}
where $\Omega = \sqrt{4\chi^2 - \gamma^2}$. The system frequency
$\Omega$ depends on the dephasing rate but we can assume that
$\gamma/2\chi \ll 1$. This is justified by the fact that in parameter
estimation, in general, it is desirable to have the parameter
dominating over the noise. Comparing the deviation
$\delta\chi_\text{c}$ to the deviation of setup (I) with maximally
entangled states,
\begin{equation}\label{eq:deviation-max}
  \delta\chi_\text{m}(t) = \frac{\eu^{N\gamma t} \sqrt{1 - \eu^{-2N\gamma t}
      \cos^2(N\chi t)}}{\sqrt{Tt}N|\sin(N\chi t)|},
\end{equation}
with $N=2$, we see that the dephasing rate is reduced by a factor of
$2$.  Furthermore, the additional term in $\langle
M_\text{c}\rangle(t)$ is of order $\gamma/\Omega$, which is given by
\begin{equation*}
  \frac{\gamma}{\Omega} = \frac{\gamma}{2\chi} \left[1 +
    \frac{\gamma^2}{8\chi^2} + \mathcal{O}\left(\left( \frac{\gamma}{\chi}
      \right)^4\right) \right].
\end{equation*}
Since $\gamma/2\chi \ll 1$, the additional terms in
Eq.~\eqref{eq:delta_chi_2} are a small perturbation compared to the
solution with maximally entangled states. Similarly, the frequencies
in the reference system and the cluster state setup are the same to
lowest order in $\gamma/2\chi$.

We calculate the minimum of $\delta\chi_\text{c}(t)$ by first finding
the envelope function of $\delta\chi_\text{c}(t)$, which is given by
the solution of $\cos(\Phi) + \alpha \sin(\Phi) = 0$. We assume that
$\alpha := \gamma/\Omega$ and $\Phi := \Omega t$ can be varied
independently. Together with the equality $\cot(x-m\pi) = \cot(x)$,
where $m$ is an integer, this yields
\begin{equation}\label{eq:cond_Omega_t}
  \Omega t = \cot^{-1}\left(-\alpha\right) = \frac{k\pi}{2} +
  \frac{\gamma}{\Omega} + \mathcal{O}\left(\left( \frac{\gamma}{\Omega}
    \right)^3\right),\quad k\text{ odd}.
\end{equation}
Plugging this into Eq.~\eqref{eq:delta_chi_2} results in the envelope
of $\delta\chi_\text{c}(t)$
\begin{equation}\label{eq:env-ana}
  \delta\chi_\text{c,env}(t) = \frac{\eu^{\gamma t}\Omega^2 \sqrt t}{2\sqrt{T}
    (\gamma + 4\chi^2 t)}.
\end{equation}
In the next step, we calculate the minimum of this envelope with
respect to $t$, which is given by
\begin{equation*}
  t_\text{c,min} = \frac{1}{4\gamma} \left(1 + \sqrt{1 -
      \frac{3\gamma^2}{\chi^2} + \frac{\gamma^4}{4\chi^4}} \right) -
  \frac{\gamma}{8\chi^2}.
\end{equation*}
Since we have $\gamma/2\chi \ll 1$, this minimal time to lowest order
is given by $t_\text{c,min} \approx 1/2\gamma - \gamma/8\chi^2$, which
is approximately 2 times the minimal time of setup (I) with maximally
entangled states~\cite{HueMacPel97} if we further assume that
$1/2\gamma \gg \gamma/8\chi^2$. Hence, in the cluster state setup the
evolution times can be longer than with maximally entangled states. It
turns out that the minimal deviation
$\delta\chi_\text{c,env}(t_\text{c,min})$ is lower than the one for
the standard scheme (I) with maximally entangled (or uncorrelated)
states. If the condition, Eq.~\eqref{eq:cond_Omega_t}, is not
satisfied exactly for a particular set of parameters, then the actual
lobes of the deviation still lie on the envelope but the minimum of
these lobes does not coincide with the minimum of the envelope making
the actual minimum of the deviation slightly larger.

We are now able to calculate the relative improvement of this scheme
with respect to the reference systems. The improvement is quantified
by
\begin{equation}\label{eq:epsilon-def}
  \epsilon(\gamma) = 1 -
  \frac{\delta\chi_\text{c,env}(t_\text{c,min})}{\delta\chi_\text{ref}},
\end{equation}
where $\delta\chi_\text{ref}$ is the minimum of the envelope of the
reference deviation. With this definition an improvement of the
cluster state scheme is given by positive values of $\epsilon$,
whereas negative values indicate that the reference system performs
better. For both maximally entangled and uncorrelated states in setup
(I) we have $\delta\chi_\text{ref} = \delta\chi_\text{m,min} =
\delta\chi_\text{u,min} = \sqrt{2\gamma \eu/NT}$.  The expansion of
$\epsilon(\gamma)$ in terms of the small parameter $\gamma/\chi$ is
given by
\begin{equation}\label{eq:epsilon}
  \epsilon(\gamma) = \epsilon\left(\frac{\gamma}{\chi}\right) =
  1-\frac{1}{\sqrt 2} + \frac{3}{4\sqrt 2} \left(\frac{\gamma}{\chi}\right)^2 +
  \mathcal{O}\left( \left(\frac{\gamma}{\chi}\right)^4 \right).
\end{equation}
In the limit $\gamma = 0$ this series assumes an offset $1-1/\sqrt{2}
\approx 0.293$. In contrast, in Sec.~\ref{sec:parameter-estimation} we
have seen that the cluster state setup and the conventional setup (I)
yield the same result for $\gamma = 0$. This discrepancy is owing to
the fact that $t_\text{c,min}$ diverges as $\gamma \rightarrow 0$. For
nonzero $\gamma$ we always find times for which Eq.~\eqref{eq:epsilon}
holds, although for small times the two deviations can lie arbitrarily
close together. The position of the minimum of the deviations
increases with $1/\gamma$, so as $\gamma$ decreases, the duration of
the experiment quickly exceeds experimentally reasonable time
scales. If we assume a fixed maximal evolution time, the improvement
can diverge from the constant for values of $\gamma$ smaller than 2
times the reciprocal of this evolution time.

\subsection{Numerical results for $N>2$}
In principle, the same analytical methods outlined above could be
applied to $N>2$. Furthermore, the analytical results could be
compared to an analytical solution for reference setup (II), which
would lead to qualitatively similar results. For simplicity, in the
following we present numerical results of the cluster state setup and
reference system (I) for higher $N$, and explain the findings with the
insight we gained from the analytical treatment of the case $N=2$. Our
numerical results for $N=2$ agree with the analytical solution above
and known analytical solutions for arbitrary $N$ with maximally
entangled and uncorrelated states so we expect a good accuracy for the
case $N>2$. Similar to the case $N=2$ we find that the cluster state
setup beats maximally entangled states in terms of lowest deviation
even for $N>2$. In our calculations we limit the integration times to
an $N$-dependent time $t_f(N)$. This choice illustrates the effect of
a finite evolution time in experiments.

\begin{figure}
  \centering
  \includegraphics[width=\linewidth]{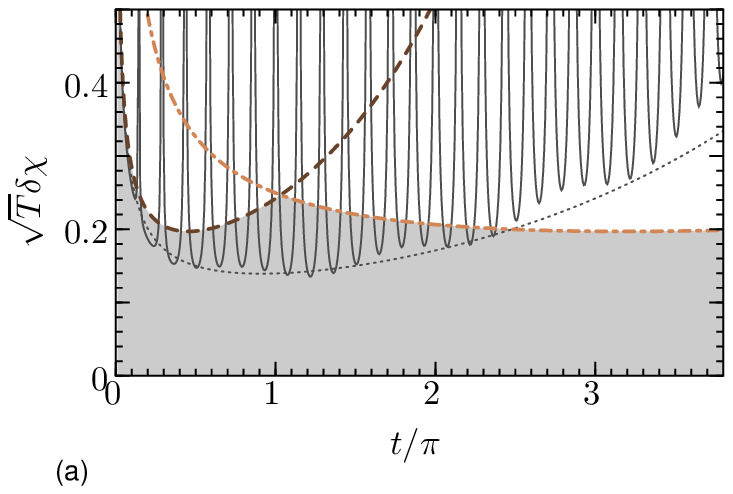}
  \includegraphics[width=\linewidth]{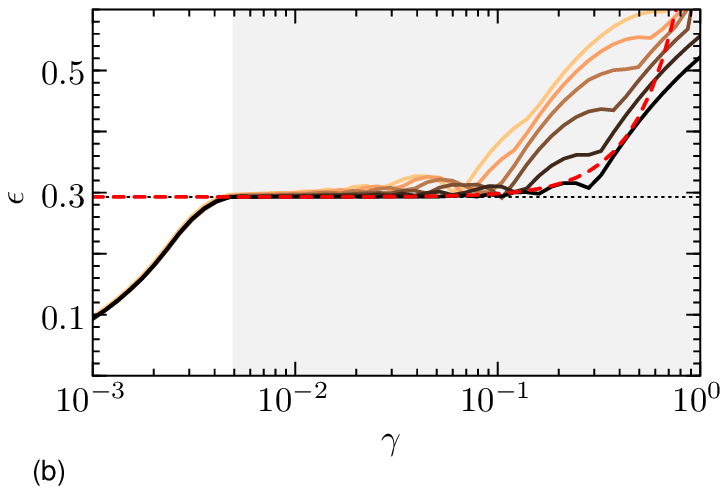}
  \caption{(Color online) Cluster states vs reference system (I).
    Plot (a) shows the rescaled deviation for the cluster state setup
    (solid line), and the envelopes for maximally entangled states,
    Eq.~\eqref{eq:env-max} (dashed line), uncorrelated states,
    Eq.~\eqref{eq:env-unc} (dashed-dotted line) in reference system
    (I), and cluster states, Eq.~\eqref{eq:env-cluster} (dotted line),
    all with $N=7$. The shaded area is the area of possible
    improvement, which is penetrated by the cluster state scheme
    deviation. In (b) we plot the improvement of the cluster state
    scheme vs maximally entangled states in reference system (I) for
    $N = 2,3,4,5,6,7$ (dark to light lines). Uncorrelated states lead
    to very similar curves. The dashed line indicates the analytical
    approximation, Eq.~\eqref{eq:epsilon}. The evolution time was kept
    constant for each $N$ at $t_f(N) = \lceil 1/0.005 N\rceil$ so we
    expect major improvements for $\gamma>0.005$ in the shaded
    area. The horizontal dotted line marks the constant improvement
    $1-1/\sqrt{2}$. For all curves $\chi = 1$ and in (a) $\gamma =
    0.05$.}\label{fig:comp-sysI}
\end{figure}

The results in Fig.~\ref{fig:comp-sysI}(a) show the deviations for the
$N=7$ cluster state setup compared to maximally entangled and
uncorrelated states in reference setup (I). The envelopes of the
deviation for maximally entangled and uncorrelated states are given by
\begin{align}
  \delta\chi_\mathrm{m,env} &= \frac{\eu^{N\gamma
      t}}{N\sqrt{Tt}},\label{eq:env-max}\\
  \delta\chi_\mathrm{u,env} &= \frac{\eu^{\gamma
      t}}{\sqrt{NTt}},\label{eq:env-unc}
\end{align}
respectively. Note that both reference systems attain the same minimum
at different times. The cluster state setup, however, attains a lower
minimum than both, which is indicated by the black curve entering the
shaded area in Fig.~\ref{fig:comp-sysI}(a). The decay rate of the
cluster state setup is approximately one-half the rate of maximally
entangled states, which allows for a longer evolution time per
experiment just as in the $N=2$ case.

The numerical analysis indicates that the envelope of the cluster
state setup can be approximately expressed as
\begin{equation}\label{eq:env-cluster}
  \delta\chi_\text{c,env} \approx \frac{\eu^{N\gamma t/2}}{N\sqrt{Tt}}.
\end{equation}
This can also be extrapolated from the analytical envelope for $N=2$,
Eq.~\eqref{eq:env-ana}. To this end, in Eq.~\eqref{eq:env-ana} we
introduce the substitutions $2\chi\rightarrow N\chi$ and
$\gamma\rightarrow N\gamma/2$. If we now assume $\gamma/2N\chi^2 \ll
1$, Eq.~\eqref{eq:env-cluster} follows. This approximation does not
take into account the small oscillations we observe for
$\delta\chi_\text{c}$ around this envelope in
Fig.~\ref{fig:comp-sysI}(a). The minimum of Eq.~\eqref{eq:env-cluster}
and the time at which the minimum is attained are given by
\begin{align}
  \delta\chi_\text{c,min} &= \sqrt{\frac{\gamma\eu}{T N}},\label{eq:min-cluster}\\
  t_\text{c,min} &= \frac{1}{N\gamma}.
\end{align}
This indicates that, for $\gamma/\chi \ll 1$, we can expect the
improvement, Eq.~\eqref{eq:epsilon-def}, of the minimal deviation to be
$\epsilon \approx 1-1/\sqrt{2} \approx 29.3\%$ compared to more common
schemes with maximally entangled and uncorrelated states. This is
consistent with our result for $N=2$, where we obtained the same
improvement to lowest order in $\gamma/\chi$. Indeed if we plot the
improvement for varying $\gamma$ for $N\ge 2$, we recover this
behavior for small $\gamma/\chi$ (onset of the light shaded area in
Fig.~\ref{fig:comp-sysI}(b)). For Fig.~\ref{fig:comp-sysI}(b) we fixed
the evolution time to $t = t_f(N)$. This means that from a finite
$\gamma_f \propto 1/N t_f(N)$ the minimum is expected to lie outside
the integration domain, which leads to the observed deviation from the
constant value in the white area ($\gamma < \gamma_f$) of the figure.
For $N=2$ we know that the improvement will deviate from the constant
value $1-1/\sqrt{2}$ as $\gamma/\chi \rightarrow 1$.

We now focus on the substructure of kinks and humps of the $N=2$ curve
(black line) in Fig.~\ref{fig:comp-sysI}(b). Analyzing the position of
the local maxima (humps) allows us to determine the frequency $\Omega
= \sqrt{4\chi^2 - \gamma^2}$. The humps are the result of the minimum
of a lobe of the actual curve for $\delta\chi$ hitting the global
minimum of the envelope of $\delta\chi_\text{c}$. This happens when
the condition, Eq.~\eqref{eq:cond_Omega_t}, is satisfied at
$t_\text{min} = 1/N\gamma$. Hence for $N=2$ we rewrite
Eq.~\eqref{eq:cond_Omega_t} as
\begin{equation*}
  \begin{split}
    \Omega t_\text{min} &= \frac{\Omega}{2\gamma} \approx \frac{k\pi}{2}\\
    &\quad\Rightarrow \gamma = \frac{2\chi}{\sqrt{(k\pi)^2 + 1}}
    \sim \frac{\chi}{k\pi/2}
    \quad\text{($k$ odd)}.
  \end{split}
\end{equation*}
For the parameters in Fig.~\ref{fig:comp-sysI}(b) the zeroth-order
approximation we used in this calculation locates the humps reasonably
accurate for $k \ge 3$. For $k=1$ one would have to take into account
higher-order terms in $\gamma/\Omega$
(cf. Eq.~\eqref{eq:cond_Omega_t}).  In between the humps for different
$k$ the kinks indicate where the lobe of $\delta\chi$ closest to the
global minimum of the envelope is least optimal. We observe that
qualitatively the same substructure persists for higher $N$ but the
humps slightly shift position and larger improvement sets in at lower
$\gamma$.

\section{Depolarization and pure damping\label{sec:depolarizing}}
In this section, we focus on two further noise channels, namely,
depolarization and pure damping with the corresponding master
equations
\begin{align}
  \frac{\di\rho}{\di t}(t) &= \im[\rho, H_\chi] + \frac{\gamma}{4}
  \sum_{j=1}^N \biggl(2\sigma_-^{(j)}\rho\sigma_+^{(j)} -
  \sigma_+^{(j)}\sigma_-^{(j)}\rho
  - \rho\sigma_+^{(j)}\sigma_-^{(j)}\notag\\
  &\quad + 2\sigma_+^{(j)}\rho\sigma_-^{(j)} -
  \sigma_-^{(j)}\sigma_+^{(j)}\rho - \rho\sigma_-^{(j)}\sigma_+^{(j)}\notag\\
  &\quad + \sigma_z^{(j)}\rho\sigma_z^{(j)} - \rho
  \biggr),\label{eq:depolarizing}\\
  \frac{\di\rho}{\di t}(t) &= \im[\rho, H_\chi]\notag\\
  &\quad + \frac{\gamma}{2} \sum_{j=1}^N
  \biggl(2\sigma_+^{(j)}\rho\sigma_-^{(j)} -
  \sigma_-^{(j)}\sigma_+^{(j)}\rho - \rho\sigma_-^{(j)}\sigma_+^{(j)}
  \biggr),\label{eq:damping}
\end{align}
respectively~\cite{BriEng93,HeiDueBri05}. Here we used $\sigma_\pm :=
(1/2) (\sigma_x \pm \im\sigma_y)$. In this section, we use system (I)
as reference.  Depolarization occurs when the system interacts with a
bath in the high-temperature limit $T \rightarrow \infty$. This drives
the qubits into a completely depolarized state, i.e.,
$\lim_{t\rightarrow\infty}\langle\sigma_z\rangle = 0$. Pure damping,
on the other hand, describes the decay of every qubit into the state
$\ket{0}$. Under the depolarizing channel,
Eq.~\eqref{eq:depolarizing}, both maximally entangled and uncorrelated
states yield the same deviation as with dephasing (assuming the same
measurements $M_\text{u}$ and $M_\text{c}$, respectively). This is the
result of their measurements only extracting the ``transversal''
evolution depending on $\sigma_x$, whereas the depolarizing channel
changes decoherence in the longitudinal direction. On the other hand,
the damping channel, Eq.~\eqref{eq:damping}, has the transversal
effect of reducing the decay rate by one-half for maximally entangled
and uncorrelated states~\cite{HeiDueBri05}. Hence for the polarizing
channel we can use envelopes, Eqs.~\eqref{eq:env-max} and
\eqref{eq:env-unc}, and for the damping channel we replace $\gamma$ by
$\gamma/2$ in these envelopes. This leads to the minima
\begin{align*}
  \delta\chi_\text{m,env} &= \sqrt{\frac{\gamma\eu}{TN}}&\text{at }
  t_\text{m,min} &= \frac{1}{N\gamma},\\
  \delta\chi_\text{u,min} &= \sqrt{\frac{\gamma\eu}{TN}}&\text{at }
  t_\text{u,min} &= \frac{1}{\gamma}.
\end{align*}
Both minima are the same and their value is, in fact, the same minimum
we approximated for cluster states in the dephasing channel in the
preceeding section. However, a priori it is unclear how the cluster
states will evolve under these different master equations.

\begin{figure}
  \centering
  \includegraphics[width=\linewidth]{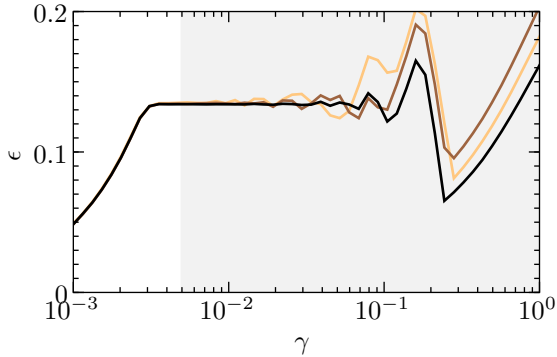}%
  \caption{(Color online) Improvement of cluster states vs maximally
    entangled states in reference system (I) under depolarization. In
    this plot $N=2, 4, 6$ (dark to light lines) and $\chi=1$. For
    uncorrelated states we get qualitatively similar results because
    both attain the same minimum. The shaded area indicates the region
    beyond the cutoff $\gamma_f = 0.005$.}\label{fig:depol_I}
\end{figure}

First we discuss numerical results for the depolarizing channel.  In
Fig.~\ref{fig:depol_I} we plot the improvement of cluster states vs
maximally entangled states in reference system (I). Similar to
dephasing noise the improvement saturates for intermediate $\gamma$
before it starts to oscillate as the noise $\gamma$ approaches the
strength of the parameter $\chi$. The constant improvement in the
depolarization case is lower than for dephasing at $\epsilon \approx
13\%$. We observe a very similar behavior if we take uncorrelated
states as reference because their deviation attains the same minimum
under depolarization as maximally entangled states. As expected, the
improvement decreases for $\gamma$ below the cutoff $\gamma_f$, which
defines a maximal evolution time as before.

For the damping channel we could not observe a global improvement of
the cluster state setup compared to either maximally entangled or
uncorrelated states in reference system (I). The deviation for cluster
states is nearly identical to the one for maximally entangled states.
This is caused by the reduced decoherence rate of both reference
states under damping as mentioned above. Hence, in a system which is
purely subject to damping, cluster states offer no improvement in
terms of minimal deviation over preparing a system in reference
(I). However, experiments are often prone to a combination of errors
and all other noise models we studied offered an overall improvement.

\section{Results for reference sytem (II)}\label{sec:system-II}

\begin{figure}
  \centering
  \includegraphics[width=\linewidth]{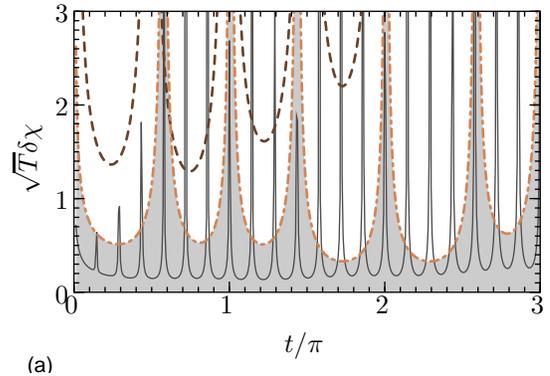}
  \includegraphics[width=\linewidth]{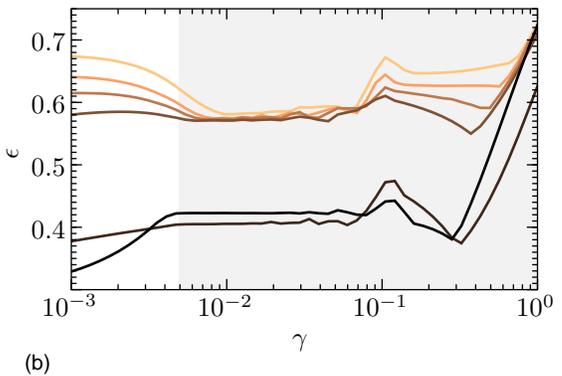}
  \caption{(Color online) Cluster state vs reference system (II) under
    dephasing noise. (a) shows the rescaled deviations and (b) the
    relative improvement with uncorrelated states. The same parameters
    and notation as in Fig.~\ref{fig:comp-sysI} apply but here we
    expose maximally entangled and uncorrelated states to reference
    system (II). In contrast to Fig.~\ref{fig:comp-sysI}, we do not
    plot envelopes in (a) but the actual deviation for the given
    parameters.}\label{fig:comp-sysII}
\end{figure}

In this section, we discuss all three noise models but now compare the
cluster state setup with reference system (II). We start with the
results for dephasing noise. In Fig.~\ref{fig:comp-sysII}(a) we plot
the deviations for maximally entangled states (dashed lines) and
uncorrelated states (dashed-dotted) and compare it to
$\delta\chi_\text{c}$ of the cluster states (solid). We do not use the
envelope as before because we did not derive an analytical solution
for the reference states in system (II). Instead the curves represent
the actual numerical deviations for a specific set of
parameters. Similarly to reference system (I), cluster states attain
the lowest overall deviation. Maximally entangled states prove useless
in this setup as their deviation is almost one order of magnitude
larger than for cluster states or even uncorrelated states.  Their
frequency also does not depend on $N$, which is the fact that causes
their advantage in the noise-free case of setup (I). The curve for
uncorrelated states exhibits a ``bimodal'' structure, where the lobes
follow two sets of envelopes with different dephasing rates. If we
plot the improvement $\epsilon$ in Fig.~\ref{fig:comp-sysII}(b), it
becomes apparent that the enhancement achievable with cluster states
is even more pronounced than in reference system (I). The nature of
the sharp divide between the improvements for $N=2, 3$ and higher $N$
is unclear. The reason for the larger improvement for $\gamma <
\gamma_f = 0.005$ in some curves is that the deviation of the
uncorrelated states reaches its minimum after the cluster states. In
this regime of $\gamma$ the reference system has not yet reached its
minimum at the cutoff time $t_f$, whereas the cluster state deviation
is closer to its minimum or has already reached it.  Similar to the
results with reference system (I), we find plateaus for small
$\gamma$, which indicate that we have again a constant offset as
$\gamma$ approaches $0$. For the resource sizes considered in the
plot, these constant values correspond to improvements of
approximately $40\%$ and $60\%$.

\begin{figure}
  \centering%
  \includegraphics[width=\linewidth]{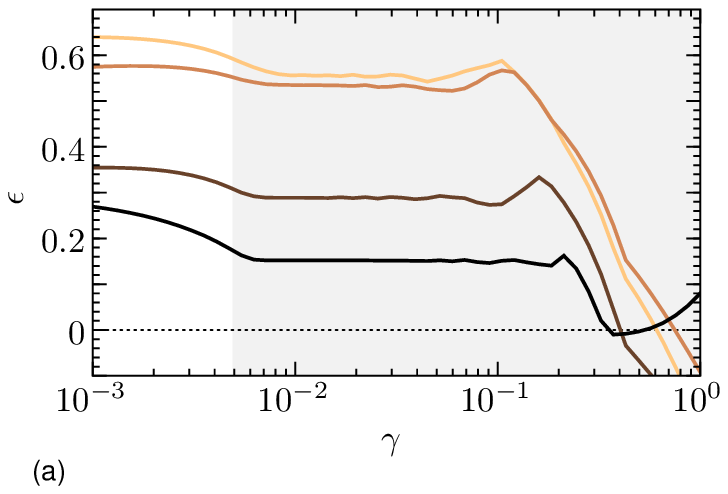}
  \includegraphics[width=\linewidth]{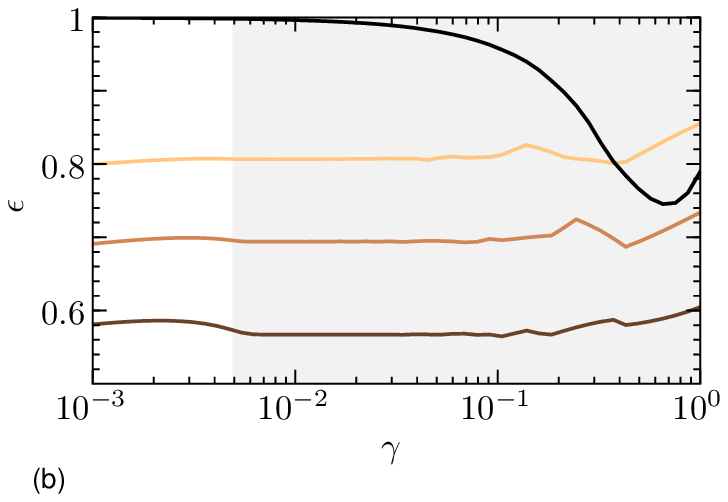}
  \caption{(Color online) Improvement of cluster states vs (a)
    uncorrelated states and (b) maximally entangled states in
    reference system (II) under damping.  Both plots show $N=2,3,4,6$
    (from dark to light) with $\chi=1$ and $\gamma_f = 0.005$ (marked
    by the shaded area).}\label{fig:depol_damp_II}
\end{figure}

In contrast to the case with system (I), when we let the reference
states evolve under the depolarizing and pure damping master
equations, we do observe a notable deviation improvement with cluster
states under these noise models.  The results for damping in
Figs.~\ref{fig:depol_damp_II}(a,b) suggest a steady increase of the
improvement for higher particle numbers. One curiosity is the result
for the $N=2$ maximally entangled state in
Fig.~\ref{fig:depol_damp_II}(b) (black curve). In this case, the
expectation value of $M_\text{m}$ does not exhibit oscillations in
time, which makes this state unsuitable for the parameter
estimation. For higher $N$ we do recover oscillations but still see a
large advantage of cluster states.  The results for depolarization are
similar with similar magnitudes of improvement for both reference
states.

\section{Scaling}\label{sec:scaling}
We have seen in Sec.~\ref{sec:parameter-estimation} that, in
principle, our scheme can achieve the Heisenberg limit in the number
of qubits $N$ for decoherence-free systems. For dephasing noise we
have given an approximation of the envelope of the deviation
Eq.~\eqref{eq:env-cluster}, which results in a shot-noise-limited
scaling, Eq.~\eqref{eq:min-cluster}.  In this section, we consider
experiments with a fixed evolution time and derive the scaling
properties of the deviation in this limit.

We consider $\gamma t$ to be a small parameter. Then we can expand
Eq.~\eqref{eq:env-cluster} to first order
\begin{equation}\label{eq:exp-cluster}
  \delta\chi_\text{c,env} = \frac{1}{N\sqrt{Tt}}
  + \frac{\gamma}{2}\sqrt{\frac{t}{T}}
  + \mathcal{O}\left(\left( N\gamma t/2 \right)^2\right).
\end{equation}
The first term on the right-hand side is a Heisenberg scaling with
$1/N$ and the second term is a constant offset of the deviation.  We
compare this expansion with the expansions of the deviation of
maximally entangled and uncorrelated states in reference system (I),
Eqs.~\eqref{eq:env-max} and \eqref{eq:env-unc},
\begin{align}
  \delta\chi_\text{m,env} &= \frac{1}{N\sqrt{Tt}}
    + \gamma\sqrt{\frac{t}{T}}
    + \mathcal{O}\left[\left( N\gamma t \right)^2\right],\label{eq:exp-max}\\
  \delta\chi_\text{u,env} &= \frac{1}{\sqrt{NTt}}
    + \frac{\gamma\sqrt{t}}{\sqrt{NT}}
    + \mathcal{O}\left[\left( \gamma t \right)^2\right].\label{eq:exp-unc}
\end{align}
First we note that the constant offset (second term on the right-hand
side) for cluster states in Eq.~\eqref{eq:exp-cluster} is only
one-half the offset for maximally entangled states in
Eq.~\eqref{eq:exp-max}, and also smaller than the second term in
Eq.~\eqref{eq:exp-unc}.  This guarantees that in this limit the
deviation of the cluster states is always smaller than in both
reference systems.  Furthermore, the approximation for cluster states
is valid for $(N\gamma t/2)^2 \ll 1$, whereas for maximally entangled
states it is only valid for $(N\gamma t)^2 \ll 1$. For fixed $\gamma
t$ we can thus expect the deviation of the cluster states to scale
with the Heisenberg limit (first term in the expansion) for 2 times as
many qubits as with maximally entangled states.  In the limit
considered here, uncorrelated states only scale with the shot-noise
limit. For small numbers of qubits and small fixed $\gamma t$ the
cluster state setup is superior in scaling to both reference schemes.


\section{Conclusion\label{sec:conclusion}}
We have introduced cluster states into parameter estimation theory. By
using ``maximally entangled'' cluster states as a probe in a system
evolving under a three-body Hamiltonian, we could show that it is
possible to achieve the Heisenberg limit of sensitivity to the
parameter. The addition of dephasing noise leads to decoherence in the
system but our results indicate that the decoherence rate is reduced
compared to the standard setup in quantum parameter estimation with
maximally entangled states. This leads to an improvement of the
minimally achievable deviation. This improvement also persists when
the cluster state setup is compared to different reference setups. The
results suggest that the improvement for a given number of qubits is
almost constant when the strength of the noise is much smaller than
the parameter, which is often desirable in experiments. Similar
results hold when the qubits are subject to depolarization or damping
noises.

The creation of 1D cluster states in optical lattices typically
results in cluster sizes on the order of $N\approx
40$~\cite{ManGreWid03}.  Also proposals for implementing three-body
Hamiltonians have been devised for optical
lattices~\cite{PacPle04,BueMicZol07}, which makes this system an ideal
candidate for our measurement scheme. The achievable number of qubits
could be increased even further by extending the present scheme to
higher-dimensional cluster states. These can be created in optical
lattices by using state-dependent lattice shifts.

Analogous to parameter estimation with maximally entangled states, the
Heisenberg scaling does not persist for large systems under
decoherence. However, for small systems we have shown that the
deviation of the cluster state setup scales at the Heisenberg
limit. In the cluster state setup, this limit can be realized for 2
times as many qubits as with maximally entangled or uncorrelated
states.

\begin{acknowledgments}
  This research was supported by the European Commission under the
  Marie Curie Programme through \caps{QIPEST}.
\end{acknowledgments}

\appendix*
\section{Analytical solution of the dephasing master equation for
  $N=2$}\label{app:solution}
To illustrate the behavior of the deviation $\delta\chi_\text{c}$, we
derive an analytical solution of the master equation~\eqref{eq:master}
with the cluster Hamiltonian Eq.~\eqref{eq:H-cluster} for the case
$N=2$. Finding a solution for higher $N$ is complicated by the fact
that the basis of the underlying Hilbert space grows exponentially
with $N$. In the following derivation we use the convention $\sigma_0
:= I$, $\sigma_{1,2,3} := \sigma_{x,y,z}$, where $I$ is the
identity. With this notation the full Hamiltonian is given by
\begin{equation*}
  H_{II} = \frac{\chi}{2} \left(K^{(1)} + K^{(2)}\right) = \frac{\chi}{2}
  \left(\sigma_1^{(1)}\sigma_3^{(2)} + \sigma_3^{(1)}\sigma_1^{(2)}\right).
\end{equation*}
The master equation takes the form
\begin{equation}\label{eq:mastern2}
  \begin{split}
    \frac{\di\rho}{\di t}(t) &= \im\frac{\chi}{2} \left\{\left[\rho(t),
	K^{(1)}\right] + \left[\rho(t), K^{(2)}\right]\right\}\\
    &\quad + \frac{\gamma}{2}\left[\sigma_3^{(1)}\rho(t)\sigma_3^{(1)}
      + \sigma_3^{(2)}\rho(t)\sigma_3^{(2)} - 2\rho(t) \right].
  \end{split}
\end{equation}
We expand $\rho(t) \in L(\mathcal{H}^{\otimes 2})$ in the Pauli basis
for two qubits
\begin{equation*}
  \rho(t) = \sum_{i,j=0}^3 c_{ij}(t) \sigma_i\otimes\sigma_j,
\end{equation*}
where $c_{ij}(t)$ are the time-dependent coefficients of the
corresponding basis vectors in $L(\mathcal{H}^{\otimes 2})$. For
$\rho(t)$ to be of unit trace we must have $c_{00}(t) = 1/2^N =
1/4$. For the commutators in Eq.~\eqref{eq:mastern2} we only need to
consider terms in the expansion of $\rho(t)$ where both Pauli matrices
in the direct product do not commute with $\sigma_1\otimes\sigma_3$
and $\sigma_3\otimes\sigma_1$. For the commutator with $K^{(1)}$
($K^{(2)}$) the commuting vectors are $\sigma_0\otimes\sigma_0$,
$\sigma_0\otimes\sigma_3$, $\sigma_1\otimes\sigma_0$, and
$\sigma_1\otimes\sigma_3$ ($\sigma_0\otimes\sigma_0$,
$\sigma_0\otimes\sigma_1$, $\sigma_3\otimes\sigma_0$, and
$\sigma_3\otimes\sigma_1$). Each commutator is then a linear
combination in the subspace orthogonal to the one spanned by the four
commuting vectors. The incoherent part of the master equation can be
easily evaluated by noting that $\sigma_3^{(i)}\rho(t)\sigma_3^{(i)}$
flips the signs of the components which contain $\sigma_1^{(i)}$ or
$\sigma_2^{(i)}$. These observations allow us to rewrite the master
equation as a system of coupled \caps{ODE}s in the coefficients, which
we write as a vector $c(t) = [c_{01}(t), c_{02}(t), \dots]$.  Note
that the component $c_{00}(t)$ is constant which does not affect the
other coefficients so we do not include it in this calculation. In
terms of this vector the system of \caps{ODE}s becomes
\begin{equation}\label{eq:ode}
  \frac{\di c}{\di t}(t) = A c(t).
\end{equation}
By evaluating the commutators and the incoherent part of the master
equation we find for the matrix $A$,
\begin{widetext}
  \begin{equation*}
    \mbox{$
      A = \left(\begin{array}{@{}ccccccccccccccc@{}}
          -\gamma & 0 & 0 & 0 & 0 & -\chi & 0 & 0 & 0 & 0 & 0 & 0 & 0 & 0 & 0 \\
          0 & -\gamma & 0 & 0 & \chi & 0 & 0 & 0 & 0 & 0 & 0 & 0 & 0 & 0 & -\chi \\
          0 & 0 & 0 & 0 & 0 & 0 & 0 & 0 & 0 & 0 & 0 & 0 & 0 & \chi & 0 \\
          0 & 0 & 0 & -\gamma & 0 & 0 & 0 & 0 & -\chi & 0 & 0 & 0 & 0 & 0 & 0 \\
          0 & -\chi & 0 & 0 & -2\gamma & 0 & 0 & -\chi & 0 & 0 & 0 & 0 & 0 & 0 & 0 \\
          \chi & 0 & 0 & 0 & 0 & -2\gamma & 0 & 0 & 0 & 0 & \im\chi & 0 & 0 & 0 & 0 \\
          0 & 0 & 0 & 0 & 0 & 0 & -\gamma & 0 & 0 & -\im\chi & 0 & 0 & 0 & 0 & 0 \\
          0 & 0 & 0 & 0 & \chi & 0 & 0 & -\gamma & 0 & 0 & 0 & 0 & 0 & 0 & -\chi \\
          0 & 0 & 0 & \chi & 0 & 0 & 0 & 0 & -2\gamma & 0 & 0 & 0 & 0 & \im\chi & 0 \\
          0 & 0 & 0 & 0 & 0 & 0 & -\im\chi & 0 & 0 & -2\gamma & 0 & 0 & -\im\chi & 0 & 0 \\
          0 & 0 & 0 & 0 & 0 & \im\chi & 0 & 0 & 0 & 0 & -\gamma & -\chi & 0 & 0 & 0 \\
          0 & 0 & 0 & 0 & 0 & 0 & 0 & 0 & 0 & 0 & \chi & 0 & 0 & 0 & 0 \\
          0 & 0 & 0 & 0 & 0 & 0 & 0 & 0 & 0 & -\im\chi & 0 & 0 & -\gamma & 0 & 0 \\
          0 & 0 & -\chi & 0 & 0 & 0 & 0 & 0 & \im\chi & 0 & 0 & 0 & 0 & -\gamma & 0 \\
          0 & \chi & 0 & 0 & 0 & 0 & 0 & \chi & 0 & 0 & 0 & 0 & 0 & 0 & 0 \\
        \end{array}\right).
      $}
  \end{equation*}
\end{widetext}
\noindent This matrix is not time dependent so the solution of the
\caps{ODE}~\eqref{eq:ode} is
\begin{equation*}
  c(t) = \eu^{A t} c(0),
\end{equation*}
where $c(0)$ are the components of the initial state. For the 1D
cluster state with $N=2$ the initial values are given by
\begin{equation*}
  c_{00}(0) = \frac{1}{4},\quad c_{11}(0) = -\frac{1}{4},\quad c_{22}(0) =
  \frac{1}{4},\quad c_{33}(0) = \frac{1}{4}
\end{equation*}
and all other components vanish. We only need to determine the
quantity of interest $\langle M_\text{c}\rangle_t =
\tr[M_\text{c}\rho(t)]$, which can be derived from the solution
$c(t)$. The measurement operator $M_\text{c} =
\sigma_3\otimes\sigma_3$ leaves only one component of $\rho(t)$ with
nonzero trace
\begin{equation*}
  \begin{split}
    \langle M_\text{c}\rangle_t &= \tr[c_{33}(t)\sigma_0\otimes\sigma_0] = 4 c_{33}(t)\\
    &= \eu^{-\gamma t} \left( \cos\left(\Omega t\right) +
      \frac{\gamma}{\Omega} \sin\left( \Omega t \right) \right),
  \end{split}
\end{equation*}
where $\Omega = \sqrt{4\chi^2 - \gamma^2}$. This leaves us with the
solution for the deviation
\begin{equation*}
  \begin{split}
    \delta\chi_\text{c}(t) &= \frac{\Delta M_\text{c}}{\sqrt{T/t}|\di \langle
      M_\text{c}\rangle_t/\di\chi|}\\
    &= \frac{\eu^{\gamma t} \sqrt{1 - \eu^{-2\gamma t} \left(
          \cos(\Omega t) + \frac{\gamma}{\Omega}\sin(\Omega t)
        \right)^2}}{\sqrt{Tt} \frac{4\chi}{\Omega} \left|
        \frac{\gamma}{\Omega} \cos(\Omega t) -
        \left(1+\frac{\gamma}{\Omega^2 t} \right) \sin(\Omega t)
      \right|}.
  \end{split}
\end{equation*}
We note that for $\gamma/\Omega \rightarrow 0$ this deviation
approaches the deviation for maximally entangled states with
Hamiltonian $H_I$ and a decoherence rate reduced by half
[cf. Eq.~\eqref{eq:deviation-max}].

In a similar way one derives the solution for $N=1$, which is given by
\begin{equation*}
  \delta\chi_\text{c}(t) = \frac{\eu^{\gamma t/2} \sqrt{1-\eu^{-\gamma t}
      \left(\cos(\Omega t/2) + \frac{\gamma}{\Omega}\sin(\Omega t/2) \right)^2}}{
    \sqrt{T t} \frac{2\chi}{\Omega} \left|\frac{\gamma}{\Omega} \cos(\Omega
      t/2) - \left(1 + \frac{2\gamma}{\Omega^2 t}\right) \sin(\Omega t/2) \right| }.
\end{equation*}
Comparing this form with $\delta\chi_\text{c}(t)$ suggests an obvious
extension to $N>2$ by replacing the dephasing rate with $N\gamma/2$
and the frequencies by $N\Omega/2$ in $\langle
M_\text{c}\rangle_t$. However, we could not verify numerically that
this generalized solution holds for $N>2$. The numerical solutions for
$N>2$ follow this extrapolation closely but not exactly [cf.
Fig.~\ref{fig:comp-sysI}(a)].


\end{document}